\documentclass[aps,showpacs,preprintnumbers,nofootinbibt,twocolumn]{revtex4}
\usepackage{graphicx}

\begin{document}

\title{Galactic rotation curves in modified gravity with non-minimal
coupling between matter and geometry}
\author{T. Harko}
\email{harko@hkucc.hku.hk}
\affiliation{Department of Physics and
Center for Theoretical and Computational Physics, The University
of Hong Kong, Pok Fu Lam Road, Hong Kong, P. R. China}

\begin{abstract}
We investigate the possibility that the behavior of the rotational velocities of test particles gravitating around galaxies can be explained in the framework of modified gravity models with non-minimal matter-geometry coupling. Generally, the dynamics of test particles around galaxies, as well as the corresponding mass deficit, is explained by postulating the existence of dark matter. The extra-terms in the gravitational field equations with geometry-matter coupling modify the equations of motion of test particles, and induce a supplementary gravitational interaction. Starting from the variational principle describing the particle motion in the presence of the non-minimal coupling, the expression of the tangential velocity of a test particle, moving in the vacuum
on a stable circular orbit in a spherically symmetric geometry,  is derived. The tangential velocity depends on the metric tensor components, as well as of the coupling function between matter and geometry. The Doppler velocity shifts are also obtained in terms of the coupling function.  If the tangential velocity profile is known,  the coupling term between matter and geometry can be obtained explicitly in an analytical form. The functional form of this function is obtained in two cases, for a constant tangential velocity, and for an empirical velocity profile obtained from astronomical observations, respectively. All the physical and geometrical quantities in the modified gravity model with non-minimal coupling between matter and geometry can be expressed in terms of observable/measurable parameters, like the tangential velocity, the baryonic mass of the galaxy, and the Doppler frequency shifts.  Therefore, these results open the possibility of directly testing the modified gravity models with non-minimal coupling between matter and geometry by using direct astronomical and astrophysical observations at the galactic or extra-galactic scale.
\end{abstract}


\date{\today}

\maketitle

\section{Introduction}

The rotation curves for galaxies or galaxy clusters should, according to
Newton's gravitation theory, show a Keplerian decrease with distance $r$ of
the orbital rotational speed $v_{tg}$ at the rim of the luminous matter, $%
v_{tg}^2\propto M(r)/r$, where $M(r)$ is the dynamical mass. However, one
observes instead rather flat rotation curves \cite{dm, BT08}.
Observations show that the rotational velocities increase near the center of
the galaxy and then remain nearly constant at a value of $v_{tg\infty }\sim
200-300$ km/s. This leads to a general mass profile $M(r)\approx
rv_{tg\infty }^2/G$ \cite{dm, BT08}. Consequently, the mass within a distance $%
r$ from the center of the galaxy increases linearly with $r$, even at large
distances, where very little luminous matter can be detected.

This behavior of the galactic rotation curves is explained by postulating
the existence of some dark (invisible) matter, distributed in a spherical
halo around the galaxies. The dark matter is assumed to be a cold,
pressureless medium. There are many possible candidates for dark matter, the
most popular ones being the weekly interacting massive particles (WIMP) (for
a recent review of the particle physics aspects of dark matter see %
\cite{OvWe04}). Their interaction cross section with normal baryonic
matter, while extremely small, are expected to be non-zero, and we may expect
to detect them directly.
%It has also been suggested that the dark matter in
%the Universe might be composed of superheavy particles, with mass $\geq
%10^{10}$ GeV. But observational results show the dark matter can be composed
%of superheavy particles only if these interact weakly with normal matter or
%if their mass is above $10^{15}$ GeV \cite{AlBa03}.
Scalar fields, Bose-Einstein condensates, or
long range coherent fields coupled to gravity have also been
used to model galactic dark matter %
\cite{scal}.

However, despite more than 20 years of intense experimental and
observational effort, up to now no \textit{non-gravitational} evidence for
dark matter has ever been found: no direct evidence of it, and no
annihilation radiation from it.

Therefore, it seems that the possibility that Einstein's (and the Newtonian)
gravity breaks down at the scale of galaxies cannot be excluded \textit{a
priori}. Several theoretical models, based on a modification of Newton's law
or of general relativity, have been proposed to explain the behavior of the
galactic rotation curves \cite{mod}.

A very promising way to explain the recent observational data \cite{Ri98,
PeRa03} on the acceleration of the Universe and on dark matter is to assume
that at large scales the Einstein gravity model of general relativity breaks
down, and a more general action describes the gravitational field.
Theoretical models in which the standard Einstein-Hilbert action is replaced
by an arbitrary function of the Ricci scalar $R$, first proposed in \cite
{Bu70}, have been extensively investigated lately. For a review of $f(R)$ generalized gravity models and on their physical implications see \cite{SoFa08}. The possibility that
the galactic dynamic of massive test particles can be understood without the
need for dark matter was also considered in the framework of $\ f(R)$
gravity models \cite{Cap2,Borowiec:2006qr,Mar1,Boehmer:2007kx,Bohmer:2007fh}%
.

A generalization of the $f(R)$ gravity theories was proposed in \cite
{Bertolami:2007gv} by including in the theory an explicit coupling of an
arbitrary function of the Ricci scalar $R$ with the matter Lagrangian
density $L_{mat}$. As a result of the coupling the motion of the massive
particles is non-geodesic, and an extra force, orthogonal to the
four-velocity, arises. The connections with Modified Orbital Newtonian Dynamics (MOND) and the Pioneer anomaly
were also explored, and it was suggested that the matter-geometry coupling may ne responsibly for the observed behavior of the galactic rotation curves. The model was extended to the case of the arbitrary
couplings in both geometry and matter in \cite{ha08}. The implications of
the non-minimal coupling on the stellar equilibrium were investigated in
\cite{Bertolami:2007vu}, where constraints on the coupling were also
obtained. An inequality which expresses a necessary and sufficient condition
to avoid the Dolgov-Kawasaki instability for the model was derived in \cite
{Fa07}. The relation between the model with geometry-matter coupling and
ordinary scalar-tensor gravity, or scalar-tensor theories which include
non-standard couplings between the scalar and matter was studied in \cite
{SoFa08a}. In the specific case where both the action and the coupling are
linear in $R$ the action leads to a theory of gravity which includes higher
order derivatives of the matter fields without introducing more dynamics in
the gravity sector \cite{So08}. The equivalence between a scalar theory and
the model with the non-minimal coupling of the scalar curvature and matter
was considered in \cite{BePa08}. This equivalence allows for the calculation
of the Parameterized Post-Newtonian (PPN) parameters $\beta $ and $\gamma $, which may lead to a better
understanding of the weak-field limit of $f(R)$ theories. The equations of
motion of test bodies in the nonminimal coupling model by means of a
multipole method were derived in \cite{Pu08}. The energy conditions and the
stability of the model under the Dolgov-Kawasaki criterion were studied in
\cite{Be09}.
The perturbation equation of matter on subhorizon scales in models with arbitrary matter-geometry coupling, as well as the effective gravitational constant $G_{eff}$ and two parameters $\Sigma$ and $\eta$, which along with the perturbation equation of the matter density are useful to constrain the theory from growth factor and weak lensing observations were derived in \cite{Ne09}. The age of the oldest star clusters and the primordial nucleosynthesis bounds were used in order to constrain the parameters of a toy model.
The problem of the correct definition of the matter Lagrangian of the theory and of the definition of the energy-momentum tensor, considered in \cite{BeLoPa08} and \cite{Fa09}, was solved in \cite{Ha10}. For a review of modified $f(R)$ gravity with geometry-matter
coupling see \cite{rev}.

It is the purpose of the present paper to investigate the possibility that the observed properties of the galactic rotation curves could be explained in the framework of the modified gravity theory with non-minimal coupling between matter and geometry, without postulating the existence of dark matter. As a first step in this study, starting from the variational formulation of the equations of motion, we obtain the expression of the tangential velocity of test particles in stable circular orbits. The tangential velocity is determined not only by the metric, as in standard general relativity, but it also depends on the explicit form of the coupling function between matter and geometry, as well as on its derivative with respect to the radial coordinate. In the asymptotic limit of a flat space-time, far away from the matter sources, the tangential velocity becomes a function of the coupling function only, and it does not decay to zero. Therefore the behavior of the neutral hydrogen gas clouds outside the galaxies, and their flat rotation curves, can be explained in terms of a non-minimal coupling between matter and geometry. Since the observations on the galactic rotation curves are obtained from the Doppler frequency shifts, we generalize the expression of the frequency shifts by including the effect of the matter-geometry coupling. Thus, at least in principle, the coupling function can be obtained directly from astronomical observations. Since the tangential velocity directly depends on the geometry matter-coupling, its knowledge allows the complete determination of the coupling function from the observational data. The form of the coupling function can be immediately obtained in the flat rotation curves region. By adopting an empirical law for the general form of the rotation curves, and by adopting some reasonable assumptions on the metric, the coupling function can be obtained exactly for the entire galactic space-time. Therefore, all the parameters of the modified gravity model with linear coupling between matter and geometry can be either obtained directly, or severely constrained by astronomical observations.

The present paper is organized as follows. The equations of motion in
modified gravity with linear coupling between matter and geometry, the
variational principle for the equations of motion, as well as the matter
Lagrangian, are presented in Section II. The tangential velocity of test
particles in stable circular orbits, and the corresponding Doppler frequency
shifts are derived in Section III. From the study of the galactic rotation
curves the explicit form of the matter-geometry coupling is obtained in
Section IV. We discuss and conclude our results in Section V. In the present
paper we use the Landau-Lifshitz \cite{LaLi} sign conventions and
definitions, and the natural system of units with $c=1$.

\section{Equations of motion in modified gravity with linear coupling
between matter and geometry}

By assuming a linear coupling between matter and geometry, the action for the
modified theory of gravity takes the form \cite{Bertolami:2007gv}
\begin{equation}
S=\int \left\{ \frac{1}{2}f_{1}(R)+\left[ 1+\zeta f_{2}(R)\right]
L_{mat}\right\} \sqrt{-g}\;d^{4}x~,
\end{equation}
where $f_{i}(R)$ (with $i=1,2$) are arbitrary functions of the Ricci scalar $%
R$, and $L_{mat}$ is the Lagrangian density corresponding to matter. The
strength of the interaction between $f_{2}(R)$ and the matter Lagrangian $%
L_{mat\text{ }}$is characterized by a coupling constant $\zeta $.

By assuming that the Lagrangian density $L_{mat}$ of the matter depends only
on the metric tensor components $g_{\mu \nu }$ only, and not on its
derivatives, the energy-momentum tensor of the matter is given by $T_{\mu
\nu }=L_{mat}g_{\mu \nu }-2\partial L_{mat}/\partial g^{\mu \nu }$. With the
help of the field equations one obtains for the covariant divergence of the
energy-momentum tensor the equation \cite{Bertolami:2007gv, Ha10}
\begin{equation}
\nabla ^{\mu }T_{\mu \nu }=2\left\{ \nabla ^{\mu }\ln \left[ 1+\zeta
f_{2}(R)\right] \right\} \frac{\partial L_{mat}}{\partial g^{\mu \nu }}.
\label{cons1a}
\end{equation}

In the following we will restrict our analysis to the case in which the
matter, assumed to be a perfect thermodynamic fluid, obeys a barotropic
equation of state, with the thermodynamic pressure $p$ being a function of
the rest mass density of the matter (for short: matter density) $\rho $
only, so that $p=p\left( \rho \right) $. In this case, the matter Lagrangian
density becomes an arbitrary function of $\rho $, so that $%
L_{mat}=L_{mat}\left( \rho \right) $. Then the energy-momentum tensor of the
matter is given by \cite{ha08,Ha10}
\begin{equation}
T^{\mu \nu }=\rho \frac{dL_{mat}}{d\rho }u^{\mu }u^{\nu }+\left(
L_{mat}-\rho \frac{dL_{mat}}{d\rho }\right) g^{\mu \nu },  \label{tens}
\end{equation}
where the four-velocity $u^{\mu }=dx^{\mu }/ds$ satisfies the condition $%
g^{\mu \nu }u_{\mu }u_{\nu }=1$. To obtain Eq.~(\ref{tens}) we have imposed
the condition of the conservation of the matter current, $\nabla _{\nu
}\left( \rho u^{\nu }\right) =0$, and we have used the relation $\delta \rho
=\left( 1/2\right) \rho \left( g_{\mu \nu }-u_{\mu }u_{\nu }\right) \delta
g^{\mu \nu }$. With the use of the identity $u^{\nu }\nabla _{\nu }u^{\mu
}=d^{2}x^{\mu }/ds^{2}+\Gamma _{\nu \lambda }^{\mu }u^{\nu }u^{\lambda }$,
from Eqs.~(\ref{cons1a}) and (\ref{tens}) we obtain the equation of motion of
a test particle in the modified gravity model with linear coupling between
matter and geometry as
\begin{equation}
\frac{d^{2}x^{\mu }}{ds^{2}}+\Gamma _{\nu \lambda }^{\mu }u^{\nu }u^{\lambda
}=f^{\mu },  \label{eqmot}
\end{equation}
where
\begin{equation}
f^{\mu }=-\nabla _{\nu }\ln \left\{ \left[ 1+\zeta f_{2}(R)\right] \frac{%
dL_{mat}\left( \rho \right) }{d\rho }\right\} \left( u^{\mu }u^{\nu }-g^{\mu
\nu }\right) .
\end{equation}

The extra-force $f^{\mu }$, generated due to the presence of the coupling
between matter and geometry, is perpendicular to the four-velocity, $f^{\mu
}u_{\mu }=0$. The equation of motion Eq.~(\ref{eqmot}) can be obtained from
the variational principle \cite{ha08, Ha10}
\begin{equation}
\delta S_{p}=\delta \int L_{p}ds=\delta \int \sqrt{Q}\sqrt{g_{\mu \nu
}u^{\mu }u^{\nu }}ds=0,  \label{actpart}
\end{equation}
where $S_{p}$ and $L_{p}=\sqrt{Q}\sqrt{g_{\mu \nu }u^{\mu }u^{\nu }}$ are
the action and the Lagrangian density for test particles, respectively,
and
\begin{equation}
\sqrt{Q}=\left[ 1+\zeta f_{2}(R)\right] \frac{dL_{mat}\left( \rho \right)
}{d\rho }.  \label{Q}
\end{equation}

The matter Lagrangian can be expressed as \cite{Ha10}
\begin{equation}
L_{mat}\left( \rho \right) =\rho \left[ 1+\Pi \left( \rho \right) \right]
-\int_{p_{0}}^{p}dp,  \label{Lm}
\end{equation}
where $\Pi \left( \rho \right) =\int_{p_{0}}^{p}dp/\rho $, and $p_{0}$ is an
integration constant, or a limiting pressure. The corresponding
energy-momentum tensor of the matter is given by \cite{Ha10}
\begin{equation}
T^{\mu \nu }=\left\{ \rho \left[ 1+\Phi \left( \rho \right) \right] +p\left(
\rho \right) \right\} u^{\mu }u^{\nu }-p\left( \rho \right) g^{\mu \nu },
\label{tens1}
\end{equation}
respectively, where
\begin{equation}
\Phi \left( \rho \right) =\int_{\rho _{0}}^{\rho }\frac{p}{\rho ^{2}}d\rho
=\Pi \left( \rho \right) -\frac{p\left( \rho \right) }{\rho },
\end{equation}
and with all the constant terms included in the definition of $p$.

\section{Stable circular orbits and frequency shifts in
modified gravity with linear coupling between matter and geometry}

The galactic rotation curves provide the most direct method of analyzing the
gravitational field inside a spiral galaxy. The rotation curves are obtained
by measuring the frequency shifts $z$ of the 21-cm radiation emission from
the neutral hydrogen gas clouds. Usually the astronomers report the
resulting $z$ in terms of a velocity field $v_{tg}$ \cite{dm}.

The starting point in the analysis of the motion of the hydrogen gas clouds
in modified gravity with linear coupling between matter and geometry is to
assume that gas clouds behave like test particles, moving in a static and
spherically symmetric space-time. Next, we consider two observers $O_{E}$
and $O_{\infty }$, with four-velocities $u_{E}^{\mu }$ and $u_{\infty }^{\mu
}$, respectively. Observer $O_{E}$ corresponds to the light emitter (i. e.,
to gas clouds placed at a point $P_{E}$ of the space-time), and $O_{\infty
} $ represent the detector at point $P_{\infty }$, located far from the
emitter, and that can be idealized to correspond to ''spatial infinity''.

Without loss of generality, we can assume that the gas clouds move in the
galactic plane $\theta =\pi /2$, so that $u_{E}^{\mu }=\left( \dot{t},\dot{r}%
,0,\dot{\phi}\right) _{E}$, where the dot stands for derivation with respect
to the affine parameter $s$. On the other hand, we suppose that the detector
is static (i.e., $O_{\infty }$'s four-velocity is tangent to the static
Killing field $\partial /\partial t$), and in the chosen coordinate system
its four-velocity is $u_{\infty }^{\mu }=\left( \dot{t},0,0,0\right)
_{\infty }$ \citep{Nuc01}.

The static spherically symmetric metric outside the galactic baryonic mass
distribution is given by
\begin{equation}
ds^{2}=e^{\nu \left( r\right) }dt^{2}-e^{\lambda \left( r\right)
}dr^{2}-r^{2}\left( d\theta ^{2}+\sin ^{2}\theta d\phi ^{2}\right) ,
\label{line}
\end{equation}
where the metric coefficients are functions of the radial coordinate $r$
only. The motion of a test particle in the gravitational field in modified
gravity with linear coupling between matter and geometry can be described by
the Lagrangian
\begin{equation}
L=Q\left[ e^{\nu \left( r\right) }\left( \frac{dt}{ds}\right)
^{2}-e^{\lambda \left( r\right) }\left( \frac{dr}{ds}\right)
^{2}-r^{2}\left( \frac{d\Omega }{ds}\right) ^{2}\right] ,
\end{equation}
where $d\Omega ^{2}=d\theta ^{2}+\sin ^{2}\theta d\phi ^{2}$. For $\theta
=\pi /2$, $d\Omega ^{2}=d\phi ^{2}$. From the Lagrange equations it follows
that we have two constants of motion, the energy $E$,
\begin{equation}
E=Qe^{\nu (r)}\dot{t},
\end{equation}
and the angular momentum $l$, given by
\begin{equation}
l=Qr^{2}\dot{\phi},
\end{equation}
where a dot denotes the derivative with respect to the affine parameter $s$. The condition $u^{\mu }u_{\mu }=1$ gives $1=e^{\nu \left( r\right) }\dot{t}%
^{2}-e^{\lambda (r)}\dot{r}^{2}-r^{2}\dot{\phi}^{2}$, from which, with the
use of the constants of motion, we obtain
\begin{equation}
E^{2}=Q^{2}e^{\nu +\lambda }\dot{r}^{2}+e^{\nu }\left( \frac{l^{2}}{r^{2}}%
+Q^{2}\right) .  \label{energy}
\end{equation}

This equation shows that the radial motion of the particles is the same as
that of a particle in ordinary Newtonian mechanics, with velocity $\dot{r}$,
position dependent mass $m_{eff}=2Q^{2}e^{\nu +\lambda }$, and energy $E^{2}$%
, respectively, moving in the effective potential
\begin{equation}
V_{eff}\left( r\right) =e^{\nu (r)}\left( \frac{l^{2}}{r^{2}}+Q^{2}\right) .
\end{equation}

The conditions for circular orbits $\partial V_{eff}/\partial r=0$ and $\dot{%
r}=0$ lead to
\begin{equation}\label{cons1}
l^{2}=\frac{1}{2}\frac{r^{3}Q\left( \nu ^{\prime }Q+2Q^{\prime }\right) }{%
1-r\nu ^{\prime }/2},
\end{equation}
and
\begin{equation}\label{cons2}
E^{2}=\frac{e^{\nu }Q\left( rQ^{\prime }+Q\right) }{1-r\nu ^{\prime }/2},
\end{equation}
respectively.

The line element given by Eq.~(\ref{line}) can be rewritten in terms of the
spatial components of the velocity, normalized with the speed of light,
measured by an inertial observer far from the source, as $%
ds^{2}=dt^{2}\left( 1-v^{2}\right) $, where
\begin{equation}
v^{2}=e^{-\nu }\left[ e^{\lambda }\left( \frac{dr}{dt}\right)
^{2}+r^{2}\left( \frac{d\Omega }{dt}\right) ^{2}\right] .
\end{equation}

For a stable circular orbit $dr/dt=0$, and the tangential velocity of the
test particle can be expressed as
\begin{equation}
v_{tg}^{2}=e^{-\nu }r^{2}\left( \frac{d\Omega }{dt}\right) ^{2}.
\end{equation}

In terms of the conserved quantities the angular velocity is given, for $%
\theta =\pi /2$, by
\begin{equation}
v_{tg}^{2}=\frac{e^{\nu }}{r^{2}}\frac{l^{2}}{E^{2}}.
\end{equation}

With the use of Eqs.~(\ref{cons1}) and (\ref{cons2}) we obtain
\begin{equation}
v_{tg}^{2}=\frac{1}{2}\frac{r\left( \nu ^{\prime }Q+2Q^{\prime }\right) }{%
rQ^{\prime }+Q}.  \label{vtg}
\end{equation}

Thus, the rotational velocity of the test body in modified gravity with
linear coupling between matter and geometry is determined by the metric
coefficient $\exp \left( \nu \right) $, and by the function $Q$ and its
derivative with respect to the radial coordinate $r$. In the standard general relativistic limit $\zeta =0$, $Q=1$, and we obtain $v_{tg}^2=r\nu '/2$.

The rotation curves of spiral galaxies are inferred from the red and blue
shifts of the emitted radiation by gas clouds moving in circular orbits on
both sides of the central region. The light signal travels on null geodesics
with tangent $k^{\mu }$. We may restrict $k^{\mu }$ to lie in the equatorial
plane $\theta =\pi /2$, and evaluate the frequency shift for a light signal
emitted from $O_{E}$ in circular orbit and detected by $O_{\infty }$. The
frequency shift associated to the emission and detection of the light signal
is given by
\begin{equation}
z=1-\frac{\omega _{E}}{\omega _{\infty }},
\end{equation}
where $\omega _{I}=-k_{\mu }u_{I}^{\mu }$, and the index $I$ refers to
emission ($I=E$) or detection ($I=\infty $) at the corresponding space-time
point \cite{Nuc01,La03}. Two frequency shifts, corresponding to maximum and
minimum values, are associated with light propagation in the same and
opposite direction of motion of the emitter, respectively. Such shifts are
frequency shifts of a receding or approaching gas cloud, respectively. In
terms of the tetrads $e_{(0)}=e^{-\nu /2}\partial /\partial t$, $%
e_{(1)}=e^{-\lambda /2}\partial /\partial r$, $e_{(2)}=r^{-1}\partial
/\partial \theta $, $e_{(3)}=\left( r\sin \theta \right) ^{-1}\partial
/\partial \phi $, the frequency shifts take the form \cite{Nuc01}
\begin{equation}
z_{\pm }=1-e^{\left[ \nu _{\infty }-\nu \left( r\right) \right] /2}\left(
1\mp v\right) \Gamma ,
\end{equation}
where $v=\left[ \sum_{i=1}^{3}\left( u_{(i)}/u_{(0)}\right) ^{2}\right]
^{1/2}$, with $u_{(i)}$ the components of the particle's four velocity along
the tetrad (i. e., the velocity measured by an Eulerian observer whose world
line is tangent to the static Killing field). $\Gamma =\left( 1-v^{2}\right)
^{-1/2}$ is the usual Lorentz factor, and $\exp \left( \nu _{\infty }\right)
$ is the value of $\exp \left[ \nu \left( \left( r\right) \right) \right] $
for $r\rightarrow \infty $. In the case of circular orbits in the $\theta
=\pi /2$ plane, we obtain
\begin{equation}
z_{\pm }=1-e^{\left[ \nu _{\infty }-\nu \left( r\right) \right] /2}\frac{%
1\mp \sqrt{r\left( \nu ^{\prime }Q+2Q^{\prime }\right) /2\left( rQ^{\prime
}+Q\right) }}{\sqrt{1-r\left( \nu ^{\prime }Q+2Q^{\prime }\right) /2\left(
rQ^{\prime }+Q\right) }}.
\end{equation}

It is convenient to define two other quantities, $z_{D}=\left(
z_{+}-z_{-}\right) /2$, and $z_{A}=\left( z_{+}+z_{-}\right) /2$, respectively \cite{Nuc01}. In the modified gravity model with linear coupling between matter and geometry these redshift factors are given by
\begin{equation}
z_{D}\left( r\right) =e^{\left[ \nu _{\infty }-\nu \left( r\right) \right]
/2}\frac{\sqrt{r\left( \nu ^{\prime }Q+2Q^{\prime }\right) /2\left(
rQ^{\prime }+Q\right) }}{\sqrt{1-r\left( \nu ^{\prime }Q+2Q^{\prime }\right)
/2\left( rQ^{\prime }+Q\right) }},
\end{equation}
and
\begin{equation}
z_{A}\left( r\right) =1-\frac{e^{\left[ \nu _{\infty }-\nu \left( r\right) %
\right] /2}}{\sqrt{1-r\left( \nu ^{\prime }Q+2Q^{\prime }\right) /2\left(
rQ^{\prime }+Q\right) }},
\end{equation}
respectively, which can be easily connected to the observations \citep{Nuc01}%
. $z_{A}$ and $z_{D}$ satisfy the relation $\left( z_{A}-1\right)
^{2}-z_{D}^{2}=\exp \left[ 2\left( \nu _{\infty }-\nu \left( r\right)
\right) \right] $, and thus in principle, by assuming that the metric tensor
component $\exp \left[ \nu \left( \left( r\right) \right) \right] $ is
known, $Q$ and $Q^{\prime }$ can be obtained directly from the astronomical
observations. This could provide a direct observational test of the galactic
geometry, and, implicitly, of the modified gravity models with linear
coupling between matter and geometry.

\section{Galactic rotation curves and the matter-geometry coupling}

The tangential velocity $v_{tg}$ of gas clouds moving like test
particles around the center of a galaxy is not directly measurable, but can
be inferred from the redshift $z_{\infty }$ observed at spatial infinity,
for which $1+z_{\infty }=\exp \left[ \left( \nu _{\infty }-\nu \right) /2%
\right] \left( 1\pm v_{tg}\right) /\sqrt{1-v_{tg}^{2}}$. Because of the
non-relativistic velocities of the gas clouds, with $v_{tg}\leq \left(
4/3\right) \times 10^{-3}$, we observe $v_{tg}\approx z_{\infty }$ (as the
first part of a geometric series), with the consequence that the lapse
function $\exp \left( \nu \right) $ necessarily tends at infinity to unity,
i. e., $e^{\nu }\approx e^{\nu _{\infty }}/\left( 1-v_{tg}^{2}\right)
\approx e^{\nu _{\infty }}\rightarrow 1$. The observations show that at
distances large enough from the galactic center $v_{tg}\approx $ constant %
\cite{dm}.

In the following we use this observational constraint to reconstruct the
coupling term between matter and geometry in the ''dark matter'' dominated
region, far away from the baryonic matter distribution. By assuming that $%
v_{tg}=$ constant, Eq.~(\ref{vtg}) can be written as
\begin{equation}
v_{tg}^{2}\frac{1}{rQ}\frac{d}{dr}\left( rQ\right) =\frac{\nu ^{\prime }}{2}+%
\frac{Q^{\prime }}{Q},  \label{Q0}
\end{equation}
and can be immediately integrated to give
\begin{equation}
Q(r)=\left( \frac{r}{r_{0}}\right) ^{v_{tg}^{2}/\left( 1-v_{tg}^{2}\right)
}\exp \left[ -\frac{\nu }{2\left( 1-v_{tg}^{2}\right) }\right] ,  \label{Q1}
\end{equation}
where $r_{0}$ is an arbitrary constant of integration. By assuming that the
hydrogen clouds are a pressureless dust ($p=0$) that can be characterized by
their density $\rho $ only, the Lagrangian of the matter (gas cloud) is
given by $L_{mat}\left( \rho \right) =\rho $. Therefore, from Eqs.~(\ref{Q})
and (\ref{Q1}) we obtain
\begin{eqnarray}\label{coup}
\zeta f_{2}\left( R\right) &=&\left( \frac{r}{r_{0}}\right)
^{v_{tg}^{2}/2\left( 1-v_{tg}^{2}\right) }\exp \left[ -\frac{\nu }{4\left(
1-v_{tg}^{2}\right) }\right] -1\approx   \nonumber\\
&&\frac{v_{tg}^{2}}{2\left( 1-v_{tg}^{2}\right) }\ln \frac{r}{r_{0}}-\frac{%
\nu }{4\left( 1-v_{tg}^{2}\right) }-\nonumber\\
&&\frac{v_{tg}^{2}}{8\left(
1-2v_{tg}^{2}\right) }\nu \ln \frac{r}{r_{0}}.
\end{eqnarray}

For a general velocity profile $v_{tg}=v_{tg}(r)$, the general solution of
Eq.~(\ref{Q0}) is given by
\begin{equation}
\sqrt{Q(r)}=1+\zeta f_{2}(R)=\sqrt{Q_{0}}\exp \left[ \frac{1}{2}\int \frac{%
v_{tg}^{2}(r)/r-\nu ^{\prime }/2}{1-v_{tg}^{2}(r)}dr\right] ,  \label{fin}
\end{equation}
where $Q_{0}$ is an arbitrary constant of integration.

For $v_{tg}^{2}$ we assume the simple empirical dark halo
rotational velocity law \cite{per}
\begin{equation}
v_{tg}^{2}=\frac{v_{0}^{2}x^{2}}{a^{2}+x^{2}},
\end{equation}
where $x=r/r_{opt}$, $r_{opt}$ is the optical radius containing 83$\%$ of
the galactic luminosity. The parameters $a$, the ratio of the halo core
radius and $r_{opt}$, and the terminal velocity $v_{0}$ are functions of the
galactic luminosity $L$. For spiral galaxies $a=1.5\left( L/L_{\ast }\right)
^{1/5}$ and $v_{0}^{2}=v_{opt}^{2}\left( 1-\beta _{\ast }\right) \left(
1+a^{2}\right) $, where $v_{opt}=v_{tg}\left( r_{opt}\right) $, and $\beta _{\ast
}=0.72+0.44\log _{10}\left( L/L_{\ast }\right) $, with $L_{\ast
}=10^{10.4}L_{\odot}$.

By assuming that the coupling between the neutral hydrogen clouds and the geometry
is small, $\zeta f_{2}(R)L_{mat}<<1$, and consequently does not significantly modify the
galactic geometry, the vacuum metric outside the baryonic mass distribution
with mass $M_{B}$, corresponding to $L_{mat}\approx 0$, is given by the static
spherically symmetric solution of the $f(R)$ modified gravity. By assuming,
for simplicity, that the galactic metric is given by the Schwarzschild
metric (which is also a solution of the vacuum field equations of the $f(R)$
modified gravity \cite{pun}), written as
\begin{equation}
e^{\nu }=e^{-\lambda }=1-\frac{2R_{0}}{x},
\end{equation}
where $R_{0}=GM_{B}/r_{opt}$, from Eq.~(\ref{fin}) we obtain
\begin{eqnarray}
&&1+\zeta f_{2}(R)=\exp \left[ \alpha \times{\rm arctanh}\left( \frac{\sqrt{1-v_{0}^{2}%
}}{a}x\right) \right] \times \nonumber\\
&&\left( 1-\frac{2R_{0}}{x}\right) ^{\beta }\frac{\left[
\left( 1-v_{0}^{2}\right) x^{2}+a^{2}\right] ^{\gamma }}{Q_0^{-1/2}x^{1/4}},
\end{eqnarray}
where
\begin{equation}
\alpha =-\frac{aR_{0}v_{0}^{2}}{2\sqrt{1-v_{0}^{2}}\left[ a^{2}+4\left(
1-v_{0}^{2}\right) R_{0}^{2}\right] },
\end{equation}
\begin{equation}
\beta =\frac{a^{2}-4R_{0}^{2}}{4\left[ a^{2}+4\left( 1-v_{0}^{2}\right)
R_{0}^{2}\right] },
\end{equation}
and
\begin{equation}
\gamma =-\frac{v_{0}^{2}\left[ a^{2}+6\left( 1-v_{0}^{2}\right) R_{0}^{2}%
\right] }{4\left( 1-v_{0}^{2}\right) \left[ a^{2}+4\left( 1-v_{0}^{2}\right)
R_{0}^{2}\right] },
\end{equation}
respectively. Thus the geometric part of the coupling between matter and
geometry can be completely reconstructed from the observational data on the
galactic rotation curves.

\section{Discussions and final remarks}

The galactic rotation curves and the mass distribution in clusters of
galaxies continue to pose a challenge to present day physics. One would like
to have a better understanding of some of the intriguing phenomena
associated with them, like their universality, and the very good correlation
between the amount of dark matter and the luminous matter in the galaxy. To
explain these observations, the most commonly considered models are based on
particle physics in the framework of Newtonian gravity, or of some
extensions of general relativity.

In the present paper we have considered, and further developed, an
alternative view to the dark matter problem \cite{Bertolami:2007gv}, namely, the
possibility that the galactic rotation curves and the mass discrepancy in galaxies and
clusters of galaxies can naturally be explained in gravitational models in which there is a non-minimal coupling between matter and geometry. The extra-terms in the gravitational field equations modify the equations of motion of test particles, and
induce a supplementary gravitational interaction, which can account for the
observed behavior of the galactic rotation curves. As one can see from Eq.~(%
\ref{vtg}), in the limit of large $r$, when $\nu '\rightarrow 0$ (in the case of the Schwarzschild metric $\nu '\approx 2M_B/r^2$), the tangential velocity of test particles at infinity is given by
\begin{equation}\label{vtg1}
v_{tg\infty}^{2}=\frac{r  }{%
r+Q/Q^{\prime }},
\end{equation}
which, due to the presence of the matter-gravity coupling, does not decay to zero at large distances from the galactic center,  a behavior which is
perfectly consistent with the observational data \cite{dm}, and is usually
attributed to the existence of the dark matter.

By using the simple observational fact of the constancy of the galactic
rotation curves, the matter-geometry coupling function can
be completely reconstructed, without any supplementary assumption.

If, for simplicity, we consider that the metric in the vacuum outside the
galaxy can be approximated by the Schwarzschild metric, with $\exp \left(
\nu \right) =1-2GM_{B}/r$, where $M_{B}$ is the mass of the baryonic matter
of the galaxy, then, in the limit of large $r$, we have $\nu \rightarrow 0$.
Therefore from Eq.~(\ref{coup}) we obtain
\begin{equation}
\zeta \lim_{r\rightarrow \infty }f_{2}\left( R\right) \approx \frac{%
v_{tg}^{2}}{2\left( 1-v_{tg}^{2}\right) }\ln \frac{r}{r_{0}}.
\end{equation}
If the galactic rotation velocity profiles and the galactic metric are known, the coupling function can be reconstructed exactly over the entire mass distribution of the galaxy.

One can formally associate an approximate ''dark matter'' mass profile $M_{DM}(r)$
to the tangential velocity profile, which is determined by the non-minimal
coupling between matter and geometry, and is given by
\begin{equation}\label{darkmass}
M_{DM}(r)\approx \frac{1}{2G}\frac{r^{2}\left( \nu ^{\prime }Q+2Q^{\prime
}\right) }{rQ^{\prime }+Q}.
\end{equation}

The corresponding ''dark matter'' density profile $\rho _{DM}\left( r\right) $ can be
obtained as
\begin{eqnarray}
&&\rho _{DM}\left( r\right) =\frac{1}{4\pi r^{2}}\frac{dM}{dr}=\frac{1}{4\pi G}\times \nonumber\\%
\nonumber\\
&&\left[ \frac{\nu ^{\prime }Q+2Q^{\prime }}{r\left( rQ^{\prime }+Q\right) }+%
\frac{\nu ^{\prime \prime }Q+\nu Q^{\prime }+2Q^{\prime \prime }}{2\left(
rQ^{\prime }+Q\right) }\right. -\nonumber\\
&&\left.\frac{\left( \nu ^{\prime }Q+2Q^{\prime }\right)
\left( rQ^{\prime \prime }+2Q^{\prime }\right) }{2\left( rQ^{\prime
}+Q\right) ^{2}}\right] .
\end{eqnarray}

In order to observationally constrain $M_{DM}$ and $\rho _{DM}$ we assume
that each galaxy consists of a single, pressure-supported stellar population
that is in dynamic equilibrium and traces an underlying gravitational
potential resulting from the non-minimal coupling between matter and
geometry. Further assuming spherical symmetry, the equivalent mass profile
induced by the geometry - matter coupling (the mass profile of the "dark matter" halo)
relates to the moments of the stellar distribution function via the Jeans
equation \cite{BT08}
\begin{equation}
\frac{d}{dr}\left[ \rho _{s}\left\langle v_{r}^{2}\right\rangle \right] +%
\frac{2\rho _{s}\left( r\right) \beta }{r}=-\frac{G\rho _{s}M_{DM}(r)}{r^{2}}%
,
\end{equation}%
where $\rho _{s}(r)$, $\left\langle v_{r}^{2}\right\rangle $, and $\beta
(r)=1-\left\langle v_{\theta }^{2}\right\rangle /\left\langle
v_{r}^{2}\right\rangle $ describe the three-dimensional density, radial
velocity dispersion, and orbital anisotropy of the stellar component, where $\left\langle v_{\theta }^{2}\right\rangle$ is the tangential velocity dispersion. By
assuming that the anisotropy is a constant, the Jeans equation has the
solution \cite{MaLo05}
\begin{equation}
\rho _{s}\left\langle v_{r}^{2}\right\rangle  =Gr^{-2\beta
}\int_{r}^{\infty }s^{2(1-\beta )}\rho _{s}\left( s\right) M_{DM}\left(
s\right) ds.
\end{equation}
With the use of Eq.~(\ref{darkmass}) we obtain for the stellar velocity dispersion the equation
\begin{equation}\label{integral}
\rho _{s}\left\langle v_{r}^{2}\right\rangle \approx \frac{1}{2}r^{-2\beta }\int_{r}^{\infty }s^{2(2-\beta )}\rho
_{s}\left( s\right) \frac{\nu ^{\prime }(s)Q(s)+2Q^{\prime }(s)}{sQ^{\prime
}(s)+Q(s)}ds.
\end{equation}

Projecting along the line of sight, the "dark matter" mass profile relates to two observable profiles, the projected stellar density $I(R)$, and to the stellar velocity dispersion $\sigma _p(R)$ according to the relation \cite{BT08}
\begin{equation}
\sigma _{P}^{2}(R)=\frac{2}{I(R)}\int_{R}^{\infty }\left( 1-\beta \frac{R^{2}%
}{r^{2}}\right) \frac{\rho _{s}\left\langle v_{r}^{2}\right\rangle r}{\sqrt{%
r^{2}-R^{2}}}dr.
\end{equation}

Given a projected stellar density model $I(R)$, one recovers the
three-dimensional stellar density from $\rho _{s}(r)=-(1/\pi
)\int_{r}^{\infty }\left( dI/dR\right) \left( R^{2}-r^{2}\right) ^{-1/2}dR$
\cite{BT08}. Therefore, once the stellar density profile $I(R)$, the stellar velocity dispersion $\left\langle v_{r}^{2}\right\rangle$,  and the galactic metric
 are known, with the use of the integral equation Eq.~(\ref{integral}) one can obtain the explicit form of the geometry-matter coupling function $Q$, and the equivalent mass profile induced by the non-minimal
coupling between matter and geometry.  The simplest analytic projected density profile is the Plummer profile \cite{BT08}, given by $I(R)={\cal L}\left(\pi r_{half}^2\right)^{-1}\left(1+R^2/r_{half}^2\right)^{-2}$, where ${\cal L}$ is the total luminosity, and $r_{half}$ is the projected half-light radius (the radius of the cylinder that encloses half of the total luminosity).

An important physical requirement for the circular orbits of the test
particle around galaxies is that they must be stable. Let $r_{0}$ be a
circular orbit and consider a perturbation of it of the form $r=r_{0}+\delta
$, where $\delta <<r_{0}$ \citep{La03}. Taking expansions of $V_{eff}\left(
r\right) $, $\exp \left( \nu +\lambda \right) $ and $Q^{2}(r)$ about $%
r=r_{0} $, it follows from Eq.~(\ref{energy}) that
\begin{equation}
\ddot{\delta}+\frac{1}{2}Q^{2}\left( r_{0}\right) e^{\nu \left( r_{0}\right)
+\lambda \left( r_{0}\right) }V_{eff}^{\prime \prime }\left( r_{0}\right)
\delta =0.
\end{equation}

The condition for stability of the simple circular orbits requires $%
V_{eff}^{\prime \prime }\left( r_{0}\right) >0$ \citep{La03}. This gives for the coupling function $Q$ the
constraint
\begin{eqnarray}
&&\left.\frac{d^{2}Q}{dr^{2}}\right| _{r=r_{0}}<\nonumber\\
&&\left.\left[ \nu ^{\prime \prime
}\left( \frac{l^{2}}{r^{2}}+Q^{2}\right) +\nu ^{\prime }\left( -\frac{2l^{2}%
}{r^{2}}+2QQ^{\prime }\right) -\frac{6l^{2}}{r^{4}}\right] \right|
_{r=r_{0}},\nonumber\\
\end{eqnarray}
a condition that must be satisfied at any point $r_0$ of the galactic space-time.

From observational as well as from theoretical point of view an important problem is to estimate an
upper bound for the cutoff of the constancy of the tangential velocities. If in the large $r$ limit the coupling function satisfies the condition $Q'/Q\rightarrow 0$, then $Q/Q'\rightarrow\infty$, and in this limit the tangential velocity decays to zero.  If the exact functional form of $Q$ is known, the value of $r$ at which the rotational velocity becomes zero can be accurately estimated.

In the present model all the relevant physical quantities, including the
"dark mass" associated to the coupling,  and which plays the role of the dark matter, its corresponding density profile, as well as the matter-geometry coupling function, are expressed in terms of observable
parameters - the tangential velocity, the baryonic (luminous) mass, and the Doppler frequency shifts of test particles moving around the galaxy. Therefore, this opens the possibility of directly testing the modified gravity models with non-minimal coupling between matter and geometry by using direct astronomical and astrophysical observations at the
galactic or extra-galactic scale. In this paper we have provided some basic
theoretical tools necessary for the in depth comparison of the predictions
of this model and of the observational results.

\section*{Acknowledgments}

This work is supported by an RGC grant of the government of the Hong Kong
SAR.

\end{document}